\documentclass[journal, 10pt]{IEEEtran}
\usepackage{amssymb}
\usepackage{amsmath}
\usepackage{cite}
\usepackage{mathrsfs}
\usepackage{makecell}
\usepackage{url}
\usepackage{xcolor}
\usepackage{bbm}
\usepackage{subfig}
\usepackage{cite,graphicx,amsmath,amssymb}
\usepackage{subcaption}
\usepackage{mdwtab}
\usepackage{multirow}
\usepackage{fancyhdr}
\usepackage{hyperref}
\usepackage{amsthm}

\makeatletter
\def\ScaleIfNeeded{%
	\ifdim\Gin@nat@width>\linewidth \linewidth \else \Gin@nat@width
	\fi } \makeatother

\begin{document}
	\title{Near-Field Sensing Enabled Predictive Beamforming: Fundamentals, Framework, and Opportunities}
	
	\author{Hao~Jiang, 
		Zhaolin~Wang, Yue~Liu, Hyundong~Shin, Arumugam~Nallanathan, and Yuanwei~Liu
		\thanks{H. Jiang, Z. Wang, and A. Nallanathan are with the School of Electronic Engineering and Computer Science, Queen Mary University of London, London E1 4NS, U.K. (e-mail: \{hao.jiang; zhaolin.wang; a.nallanathan\}@qmul.ac.uk).
			
			Y. Liu is with the Faculty of Applied Sciences, Macao Polytechnic
			University, Macau, SAR, China (e-mail: yue.liu@mpu.edu.mo).
			
			H. Shin is with the Department of Electronics and Information Convergence
			Engineering, Kyung Hee University, 1732 Deogyeong-daero, Giheung-gu,
			Yongin-si, Gyeonggi-do 17104, Republic of Korea (e-mail: hshin@khu.ac.kr).
			
			Y. Liu is with the Department of Electrical and Electronic Engineering, The University of Hong Kong, Hong Kong (e-mail: yuanwei@hku.hk).
		}
	}
	
	\markboth{Manuscript}
	{Manuscript}
	\maketitle
	\begin{abstract}
		The article proposes a novel near-field predictive beamforming framework for high-mobility wireless networks.
		Specifically, due to the spherical waves and non-uniform Doppler frequencies brought by the near-field region, the new ability of full-dimensional location and velocity sensing is characterized.
		Building on this foundation, the near-field predictive beamforming framework is proposed to proactively design beamformers for mobility users following arbitrary trajectories.
		Compared to the conventional far-field counterpart, the near-field predictive beamforming stands out due to: i) \emph{Prior-Knowledge-Free Prediction}, and ii) \emph{Low-Complexity and Generalizable System Design}.
		To realize these advantages, the implementation methods are discussed, followed by a case study confirming the benefits of the proposed framework. 
		Finally, the article highlights promising research opportunities inspired by the proposed framework.
	\end{abstract}
	
	\section{Introduction}
	The fast development of telecommunication technology has completely transformed people's daily lives and offers new application scenarios, such as live streaming, virtual reality, and more.  
	As the communication functionality has been exploited, the upcoming sixth generation (6G) communication network is anticipated to explore the additional sensing functionality of wireless signals \cite{liu2023survey, wild2021joint}.
	
	Driven by this technical trend, integrated sensing and communication (ISAC) has emerged as a promising candidate in recent years to fulfill this vision \cite{liu2020radar}.
	In particular, instead of treating communication and sensing as separate functionalities, ISAC enables shared use of hardware and radio resources between them, thereby reducing spectrum consumption and hardware costs.
	As a unified framework that enables both data transmission and environmental awareness, ISAC is envisioned as a key enabler of intelligent massive connectivity in 6G and beyond \cite{yuan2021integrated}.
	There are two paradigms of ISAC: i)~\emph{communication-assisted sensing}, and ii)~\emph{sensing-assisted communication}.
	In the first paradigm, communication signals reflected from sensing targets are leveraged to enable sensing functionality;
	On the contrary, the second paradigm enhances communication performance by utilizing real-time environmental information obtained through sensing.
	In practice, the latter is particularly favorable from the viewpoint of telecommunication system design, as the environmental awareness can be exploited to realize real-time low-complexity and decentralized system designs \cite{liu2022integrated}. 
		This makes sensing-assisted communication a practical and efficient approach in many dynamic scenarios.
	
	As a notable example within this category, predictive beamforming has drawn great attention from both industry and academia \cite{liu2020radar}.
	This framework utilizes ISAC signals to consistently detect the mobility statuses of moving users, thereby tracking their trajectories.
	Then, by combining the tracking results with a kinematic model, future user positions can be predicted, facilitating efficient beamforming design, i.e., steering beams towards predicted target positions.
	As such, pilots used for channel estimation (CE) can be eliminated, reducing communication overheads and saving valuable coherent time for transmissions.
	This capability is particularly beneficial for high-mobility networks such as vehicular communications.
	Moreover, due to the adoption of ISAC beams, the uplink feedback through the dedicated channel is also eliminated, thereby reducing the complexity of system design.
	However, limited by the planar-wave assumption in the far-field, conventional predictive beamforming frameworks rely on prior knowledge of trajectory shapes, known as the state evolution model, to compensate for their limited sensing capabilities.
	
	Recently, the deployment of extremely large antenna arrays (ELAAs) and the use of high-frequency transmission bands have led to pronounced near-field effects, offering a new sensing dimension that helps overcome the limitations of conventional far-field sensing.
	Specifically, the space in front of antenna arrays can be divided into the near-field and far-field regions, in a near-to-far fashion.
	The former is governed by spherical waves, while the latter is dominated by planar waves.
	The boundary between these regions is the Rayleigh distance, which can be extended to tens, even hundreds of meters \cite{lu2024hierarchial}.
	Due to the spherical waves in the near-field, two impinging signals arriving from the same angle but at different distances exhibit distinctive phase distributions, which can be exploited for the estimation of both angle and distance \cite{cui2023near}.
	Moreover, for moving targets, the Doppler frequency induced by mobility is non-uniform over the non-negligible aperture size, allowing the capture of both radial and transverse velocities \cite{giovannetti2024performance}.
	Hence, the full-dimensional mobility statuses of targets are obtainable in the near-field region, in contrast to the angle-only and radial-velocity-only far-field sensing.
	Building on the additional sensing dimensions, predictive beamforming can be enhanced to accommodate more complex scenarios.
	Motivated by these insights, we propose a novel near-field sensing-based predictive beamforming framework.
	
	\begin{figure*}[ht!]
		\centering
		\includegraphics[width=0.75\linewidth]{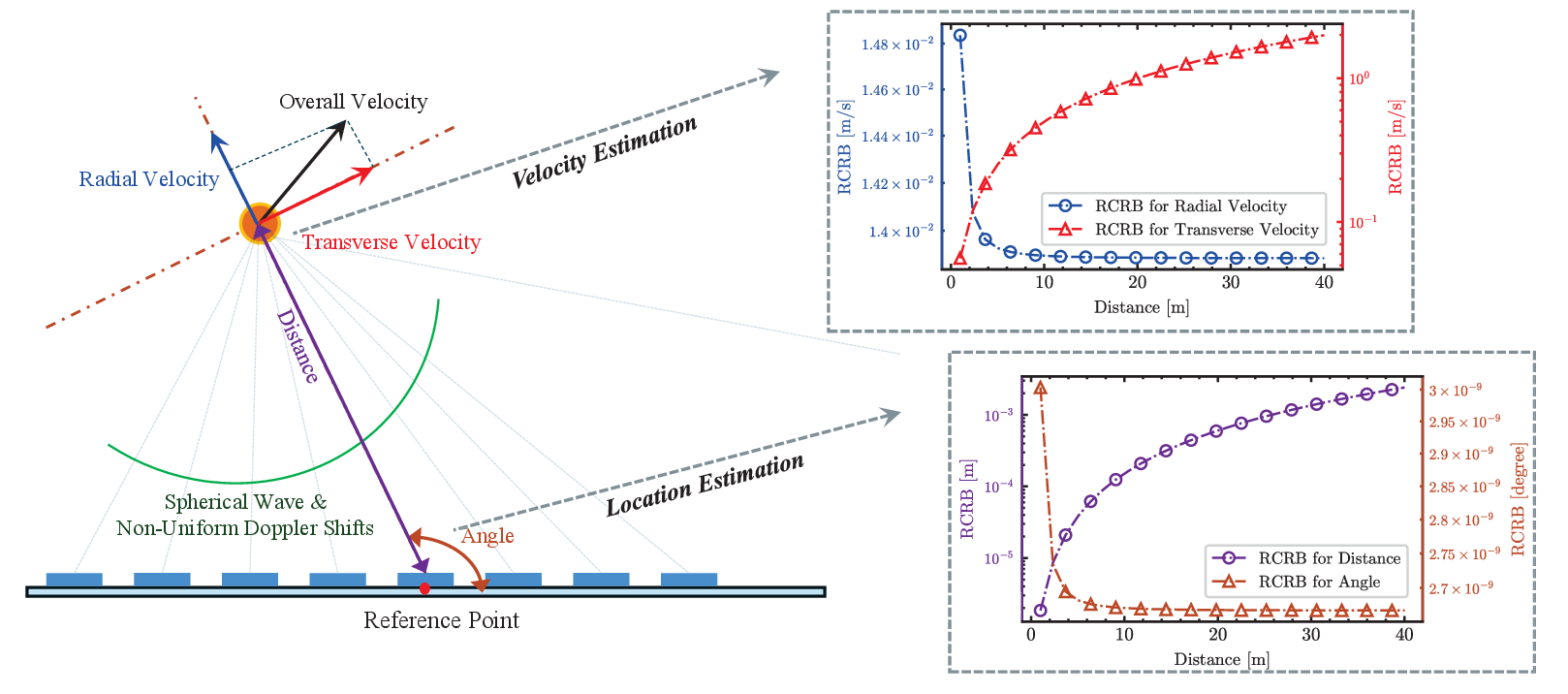}
		\caption{An illustration of the near-field and far-field sensing with RCRB stands for the root of Cramér–Rao bound. The target is placed at an angle of $\theta=\frac{\pi}{2}$ to the direction of $x$-axis and with a radial velocity $v_{r}=3$ m/s and a transverse velocity of $v_{\theta}=2$ m/s. The carrier frequency, transmit power, and the noise power are set to $28~{\mathrm{GHz}}$, $30$ dBm, and $-90$ dBm, respectively. For a fair comparison, we factor out the impact of path loss.}
		\label{fig:system_model}
	\end{figure*}
	
	\begin{table*}[h]
		\centering
		\small
		\renewcommand{\arraystretch}{1.3}
		\begin{tabular}{|c|c|c|c|c|}
			\hline
			\textbf{Category} & \textbf{Wave Type} & \textbf{Location Estimation} & \textbf{Doppler Frequency} & \textbf{Velocity Sensing}\\
			\hline
			\textbf{Far-Field Sensing} & Planar & Angle & Uniform & Radial velocity \\
			\hline
			\textbf{Near-Field Sensing} & Spherical & Angle and Distance & Non-uniform   & Radial and transverse velocities \\
			\hline
		\end{tabular}
		\caption{Comparison of far-field sensing and near-field sensing. }
		\label{tab:velocity_sensing}
	\end{table*}
	
	In this article, we first review the fundamentals behind the transition from conventional far-field partial sensing to near-field full-dimensional sensing, considering the impact of the spherical waves and non-uniform Doppler frequencies.
	In the sequel, a near-field predictive beamforming framework is proposed, detailing its principle, advantages, and implementation methods.
	Finally, we highlight some research opportunities for future endeavors, exploring potential extensions and applications of the proposed framework.

	\section{Fundamentals of Near-Field Sensing and Predictive Beamforming}
	In this section, we introduce the fundamentals of near-field sensing and predictive beamforming.
	In particular, in the first part, we elaborate on how full-dimensional sensing is achieved in the near-field region, which lies as the theoretical foundation for predictive beamforming.
		In the second part, the motivations and implementation details for predictive beamforming are explained, highlighting how sensing assists communication.

	\subsection{Fundamentals of Full-Dimensional Sensing in Near-Field}
	Compared to conventional far-field sensing, the most conspicuous characteristic of near-field is the non-negligible antenna aperture size, resulting in non-uniform phase distributions over it.
		Harnessing this feature, more comprehensive sensing is now achievable in near-field, featuring: i) \emph{Spherical-Wave-Induced Full-Dimensional Location Sensing}, enabling joint angle-distance estimation beyond the angle-only estimation in far-field, and ii)~\emph{Non-Uniform Doppler Frequency-Induced Full-Dimensional Velocity Sensing}, achieving joint radial-transverse velocity estimation beyond the radial-only velocity estimation in far-field.
		In what follows, we provide detailed narratives regarding these two features.
	
	\subsubsection{Near-Field Full-Dimensional Location Sensing}\label{sect:full-dimensional-location-sensing}
	As shown in the left-hand part of Fig. \ref{fig:system_model}, the curvatures of spherical waves in the near-field region will be preserved and lead to non-uniformly distributed phase shifts at the receiver.
		Therefore, two signals arriving from the same angle but different distances showcase different curvatures, allowing them to be distinguished.
		However, as the distance between transmitter and sensing target, i.e., user, increases, this curvature gradually diminishes and can eventually be regarded as a planar wave with negligible curvature, implying that near-field sensing degenerates into conventional far-field sensing.
		To better illustrate this, we plot the root of Cramér–Rao bound (RCRB) as a function of the distance between the transmitter and the sensing target, which is shown at the lower right part in Fig. \ref{fig:system_model}.
		It is noted that RCRB is the theoretical lower bound on the variance of any unbiased estimator, and therefore can be interpreted as a lower bound of estimation error.
		As illustrated by this figure, as the distance becomes larger, the RCRB for distance increases accordingly, while being accompanied by a correspondingly decreasing RCRB for angle.
		This observation demonstrates the degeneration from near-field sensing to far-field sensing, where the angle can be better estimated, while the distance information will eventually be lost. 
	
	\subsubsection{Near-Field Full-Dimensional Velocity Sensing}
	Similar to the stationary-target scenario in Section \ref{sect:full-dimensional-location-sensing}, the Doppler shifts caused by the mobility of the target are also non-uniform, as a result of the non-negligible aperture size of antenna arrays.
		For a clear presentation, we define the \emph{radial velocity} as the velocity component with respect to the reference point on the antenna array, and the \emph{transverse velocity} as the velocity component perpendicular to the line connecting the reference point and the target, which are geometrically showcased in the left-hand part of Fig. \ref{fig:system_model}.
		It is worth noticing that these two velocity components are the basis vectors of the space containing all possible overall velocities.
		Due to the non-negligible aperture size, the velocity components with respect to different antennas, i.e., relative velocities, are non-uniform and distinguishable, thereby causing nonidentical Doppler frequencies across the antenna array.
		Leveraging on this characteristic, both the radial and transverse velocities can be effectively captured using the velocity-projection methods in \cite{wang2025near} and \cite{jiang2024nearfieldsensingenabledpredictive
		}.
		In contrast, the difference of Doppler shifts induced by relative velocities in the conventional far-field case is too small to be detected, as the physical sizes of antenna apertures are negligible compared to the propagation distance.
		Since the differences in Doppler shifts are caused by transverse velocity, the information of this velocity component will be lost for the far-field scenarios.
	To illustrate this, the upper-right panel of Fig. \ref{fig:system_model} shows how the RCRBs of the velocity estimation evolve as the distance between the transmitter and the sensing target increases.
		As the target is away from the transmitter, it is observed that the RCRB for radial velocity decreases, while that for transverse velocity increases.
		These trends of curves show the transition from near-field velocity sensing to the far-field counterpart, meaning only the radial velocity can be accurately estimated in the far-field scenario.
	
	To sum up, we present the comparison between the far-field and near-field sensing for readers in Table \ref{tab:velocity_sensing}.

	\subsection{Fundamentals of Predictive Beamforming} \label{sect:pf}
	In high-frequency MIMO communication networks, beams are narrow and highly directional.
	Therefore, a small mismatch between the beam steering direction and the users' location may lead to severe performance loss.
	To alleviate this issue, channel state information (CSI) is essential for beamforming design, which is typically obtained through high-complex channel estimation (CE) algorithms with dedicated pilots \cite{yang2025near}. 
	Although effective, the computational complexity of these CE methods scales with the number of antennas, thus introducing intolerable signal processing overheads.
	To mitigate this issue, the beam training framework is proposed, which leverages predefined beam codebooks to identify the beam with the highest gain \cite{zhang2022fast}.
	Specifically, the codewords in the predefined codebooks are associated with beams that are steered to different angles in space.
	Thus, by traversing codebooks either exhaustively or hierarchically, the codeword, i.e., beam, offering the highest gain, can be identified.
	In this case, high-complexity signal processing techniques required by CE are eliminated and replaced by a simple searching procedure.
	
	Although the beam training framework is effective in stationary or low-mobility scenarios, its extension to the high-mobility case would be problematic.
	Particularly, the changes in the user's positions will cause rapid variations in wireless channels.
	Therefore, beam training needs to be conducted frequently to guarantee consistently reliable transmission, which will waste valuable coherent time.
	To tackle this challenge, the beam tracking framework is devised to deal with the non-stationary user scenarios.
	Instead of performing exhaustive beam searching in the beam training framework, the beam tracking framework leverages its knowledge of the current user's position and therefore searches only a subset of codewords corresponding to several adjacent positions around the user.
	Although effective for low-mobility scenarios, this approach lacks robustness against high-mobility scenarios. 
	Specifically, beam tracking operates reactively, relying on current tracking results. 
	In this case, if targets move at high velocities, outdated tracking data can degrade communication performance.
	Furthermore, as highlighted in \cite{liu2020radar}, dedicated uplink feedback is required for both the beam training and the beam tracking framework.
	\begin{figure*}[ht!]
		\centering
		\includegraphics[width=0.65\linewidth]{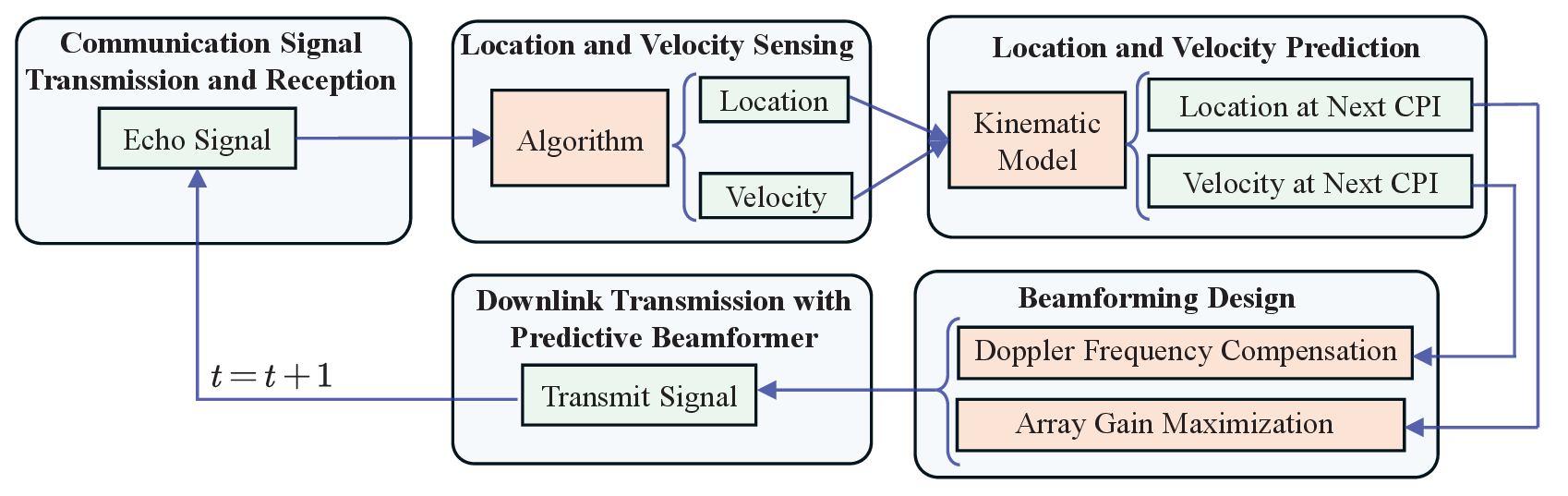}
		\caption{An illustration of the near-field predictive beamforming (NFPB) framework for a coherent processing interval (CPI).}
		\label{fig:pf_franework}
	\end{figure*}
    
	To overcome these drawbacks, the predictive beamforming framework has been proposed, which leverages sensing results to assist communication. 
	In particular, the transmitter sends ISAC beams to the target and receives the echo signal reflected from the target.
	Based on the reception of echo signals, the current location and velocity can be extracted via signal processing techniques.
	Subsequently, harnessing the sensing results, the transmitter predicts the target's position and velocity (referred to as mobility status) in the upcoming time instant, and designs the beamforming vector proactively to facilitate the reliable transmission.
	From the perspective of communication protocols, the predictive beamforming framework is composed of the following steps:  \textbf{1)~Initial Access:} Initial access is utilized for obtaining the initial velocity and position of the user.
	This stage is performed only once when the user enters the coverage of the transmitter.
	This can be achieved either by beam training or by directly reporting from the user. 
	\textbf{2)~Dataframe Transmission:} In this stage, the transmitter conveys the downlink communication dataframe to the user while simultaneously collecting reflected echo signals from the user.
	\textbf{3)~Prediction:} Based on the collected echo signals from the dataframe, the transmitter will extract the current mobility status of the user via signal processing techniques, and design beamformers in a predictive manner.
	
	As an advanced technique, the benefits of the predictive beamforming framework over the conventional beam training/tracking framework are summarized as follows:
	\begin{itemize}
		\item \textbf{Elimination of Dedicated Uplink Feedback Link:}
		By leveraging on sensing functionalities of wireless signals, the mobility status of the user in the subsequent time instant can be predicted using the information contained in received echo signals, thus eliminating the need for dedicated uplink reports from the user.
		
		\item \textbf{Reduction on Communication Overheads}:
		In either beam training or beam tracking, the whole or a subset of the predefined codebooks needs to be exhaustively swept through, resulting in communication overheads.
			In contrast, the predictive beamforming framework predicts the exact user's mobility status for the upcoming time instant, thus reducing communication overheads and saving coherent time for the upcoming transmission stage.
	\end{itemize}
	
	\section{A Near-Field Predictive Beamforming Framework}
	Building on the synergy between the full-dimensional near-field sensing and the predictive beamforming framework, a near-field predictive beamforming framework (NFPB) is proposed.
	In this section, we present the principles and advantages of NFPB first.
	Then, we introduce several implementation methods for realizing NFPB, followed by a case study on NFPB.
	For analytical simplicity, time is slotted into multiple coherent processing intervals (CPIs).
	Due to the short duration of each CPI, the target's mobility status is unchanged.

	\subsection{Principles and Advantages}\label{sect:superoirity}
	Predictive beamforming for the conventional far-field scenario has been extensively investigated in existing literature \cite{liu2020radar, liu2022learning, yuan2021bayesian}.
	However, predictive beamforming tailored for near-field scenarios is still insufficiently explored.
	Integrated within the standard protocol detailed in Section \ref{sect:pf}, the unique feature of NFPB lies in how the full-dimensional mobility status is extracted and how beamforming is designed.
		In particular, the NFPB steps are demonstrated in Fig. \ref{fig:pf_franework}, consisting of five steps:
		1)~the transmitter receives echo signal reflected by the user; 2)~the full-dimensional mobility status in current CPI, including angle $\theta$, distance $r$, radial velocity $v_{r}$, and transverse velocity $v_\theta$, are extracted by dedicated algorithms;
		3)~the position and velocity predictions for the subsequent CPI are obtained by feeding the current mobility status to the following kinematic model:
		\begin{align}
			\begin{cases}
				r^{\prime}=r+v_{\mathrm{r}}\Delta T\\
				\theta ^{\prime}=\theta +v_{\theta}\Delta T\\
			\end{cases} \notag
		\end{align}
		where $\Delta T$ is the duration for one CPI, and $(r^\prime, \theta^\prime)$ denotes the predicted polar-domain location in the next CPI;
		for the velocity predictions, $v_{\mathrm{r}}^\prime$ and $v_{\mathrm{\theta}}^\prime$ are approximated by their current values $v_{\mathrm{r}}$ and $v_{\mathrm{\theta}}$, due to the small duration of a CPI \cite{wang2025near};
		4)~beamforming based on the sensed parameters is performed, incorporating \emph{Array Gain Maximization}, aiming at focusing the beam at $(r^\prime, \theta^\prime)$, and \emph{Doppler Frequency Compensation}, aiming at compensating the frequency shits induced by velocities $(v_{\mathrm{r}}^\prime, v_{\mathrm{\theta}}^\prime)$;  
		and 5)~the overall beamforming vector is obtained and used for data transmission.
	
	The advantages of NFPB lie in the fact that NFPB overcomes a fundamental limitation of the conventional far-field predictive beamforming.
	Specifically, although far-field beam steering only requires angle information, the predictions of angle necessitate both the radial and transverse velocities according to the kinematic model.
	However, uniform Doppler frequency measurements provide only radial velocity, creating an inconsistency.
	To address this issue, conventional far-field predictive beamforming relies on prior knowledge of target trajectories to simplify the kinematic model, reducing it to a function of radial velocity alone, named as state evolution model.
	As such, the full-dimensional mobility status can be inferred from partial observations, i.e., radial velocity.
	However, this approach does not apply to some scenarios where target trajectories are too complex to be captured by analytical state evolution models, such as vehicle-to-infrastructure (V2I) networks in urban areas or unmanned aerial vehicle (UAV)-aided networks.
	
	However, in the near-field region, both the transverse and radial velocities can be inherently captured from the non-uniform Doppler frequency.
	As a result, the full-dimensional mobility status can be fully determined from the kinematic model without relying on prior knowledge.
	By leveraging the advantages of the near-field, predictive beamforming becomes more flexible and applicable to a wider range of scenarios.
	In particular, compared to the conventional far-field counterpart, the advantages of the near-field predictive beamforming are two-fold:
	\begin{itemize}
		\item \textbf{Prior-Knowledge-Free Prediction:} 
		The conventional far-field predictive beamforming frameworks necessitate the state evolution models as prior knowledge about the communication network.
		These models are derived from kinematic equations and system topologies, specifically tailored to different scenarios.
		In contrast, the proposed NFPB framework can relax this requirement by leveraging spherical waves and non-uniform Doppler frequencies.
		As a result, this framework operates without prior knowledge about user trajectories, making it more flexible.
		
		\item \textbf{Low-Complexity and Generalizable System Design}: 
		In the far-field scenarios, the state evolution model needs to be derived from the original kinematic model, based on the prior knowledge about the trajectory layout.
		This will complicate system design due to the challenge of acquiring prior knowledge, including the complexity of a sophisticated state evolution model, and the usage of wideband signals for distance acquisition.
		More importantly, far-field predictive beamforming needs to be redesigned when the scenario or state evolution model changes.
		In contrast, the full-dimensional mobility statuses of targets can be directly extracted from the spherical waves and non-uniform Doppler frequencies in the near-field, reducing complexity and enhancing generalizability.
	\end{itemize}

	\subsection{Implementation Methods} \label{sect:methods}
	Since the crucial step in NFPB lies in the estimation of mobility status from the echo signal, we present several implementation methods for readers.
	\begin{figure}[t!]
		\centering
		\includegraphics[width=0.75\linewidth]{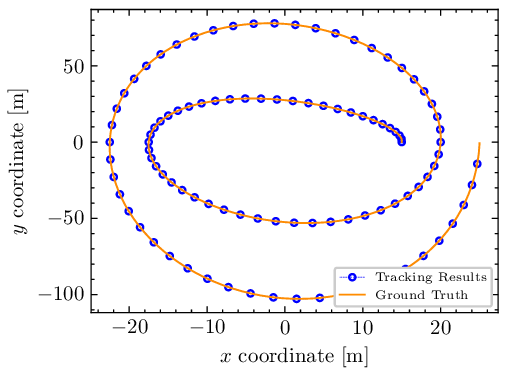}
		\caption{An illustration of location tracking results using NFPB.}
		\label{fig:trajectory}
	\end{figure}
	\subsubsection{Estimation-Based Approach}
	The estimation method directly estimates the mobility status in the current CPI based on the received echo signals.
	Subsequently, the prediction of the mobility status of the upcoming CPI is conducted by adopting the kinematic model of the target.
	In this case, in order to guarantee a reliable prediction, the estimation step is of high importance.
	The crux of this step lies in dealing with highly coupled location and velocity information in the phase of echo signals.
	The most notable method is the maximum likelihood (ML) estimation technique.
	The objective of ML estimation is to minimize the value of the likelihood function to ensure that the estimated parameters most likely match the real values.
	In addition to this method, more sophisticated methods, such as compressive sensing and Bayesian estimation, can also be utilized for mobility status estimations with lower computational complexity.
	As an advantage, estimation-based approaches are typically straightforward to implement.
	However, the drawbacks of these methods are two-fold: 1) owing to the large number of antennas on ELAAs, the echo signal vectors are high-dimensional, still incurring high computational complexity; and 2) since estimation is performed within a single CPI, these methods may suffer from reduced robustness against observation noise and error accumulation issue.
	
	\begin{figure}[t!]
		\centering
		\includegraphics[width=0.9\linewidth]{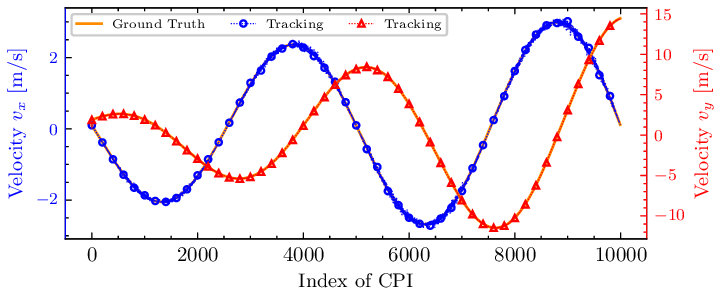}
		\caption{An illustration of velocity tracking results using NFPB.}
		\label{fig:velocity}
	\end{figure}
	\begin{figure*}[t!]
		\centering
		\includegraphics[width=0.7\linewidth]{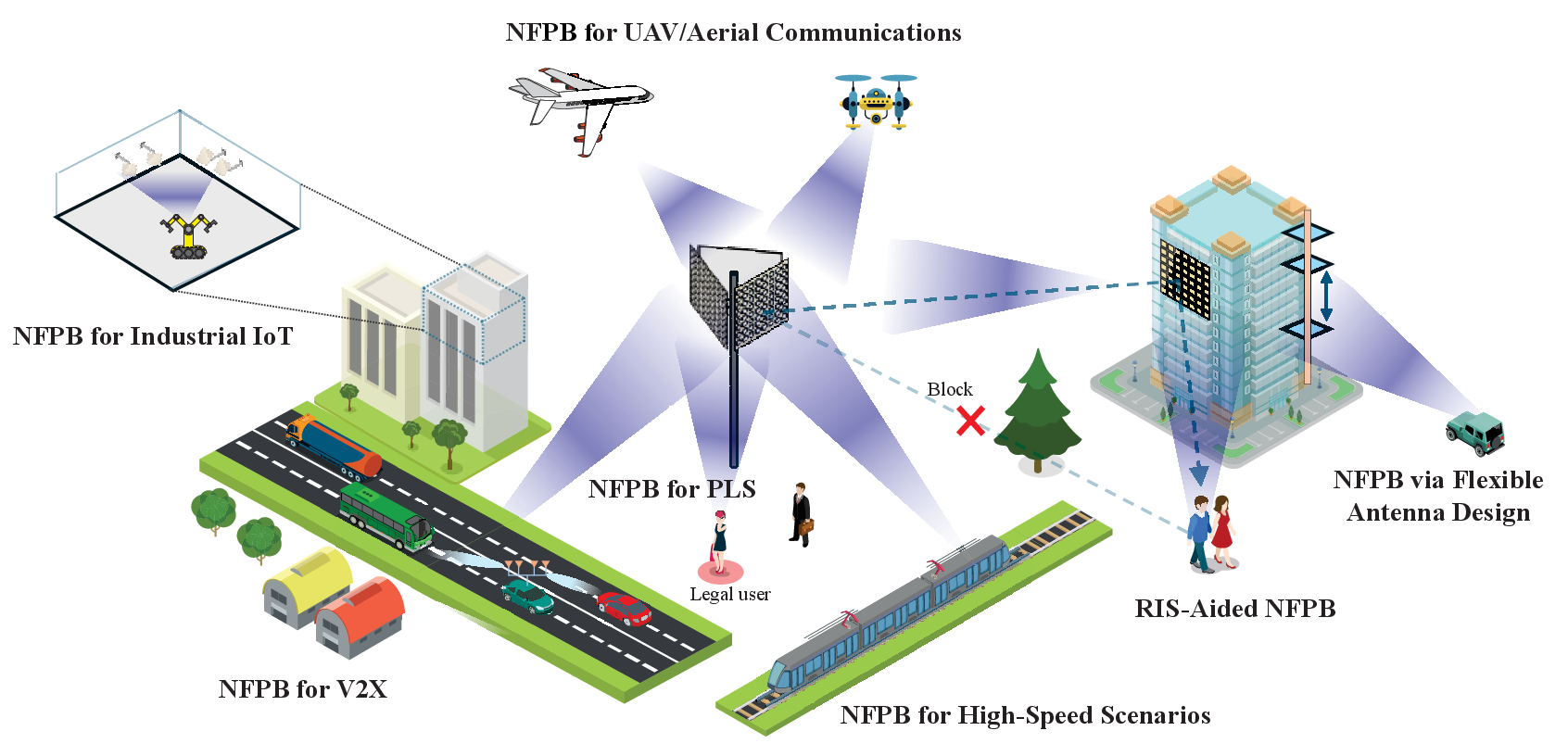}
		\caption{Illustration of research opportunities for near-field predictive beamforming framework (NFPB).}
		\label{fig:future_directions}
	\end{figure*}
	\subsubsection{Nonlinear Filter-Based Approach}
	In contrast to the estimation-based approaches that rely on echo signals from one single CPI, the filter-based approaches can exploit a sequence of observations, i.e., received echo signals from multiple CPIs, to iteratively update their prediction on the mobility status and noise covariance matrix.
	Due to the non-linear property of the observation model, the extended Kalman filter (EKF), known as a notable example in this category, utilizes Taylor expansion to linearize the observation function locally.
	Specifically, the EKF consists of two steps: \emph{Prior Prediction} and \emph{Posterior Update}.
	In the first step, the prior target mobility status and noise covariance matrix predictions are computed according to the kinematic model and its first-order derivative.
	Then, based on the received echo signals, the prior predictions will be updated according to these posterior observations.
	By iteratively performing the two steps, high-fidelity predictions can be achieved.
	It is noted that, since multiple observations are utilized for posterior updates, the Kalman filter-based method is more robust to the disturbance imposed by observation noise, as shown in \cite{jiang2024nearfieldsensingenabledpredictive}.
	However, in more complex scenarios, the EKF method may give poor performance due to the error incurred by linear approximation on highly nonlinear functions.
	Therefore, the unscented Kalman filter (UKF) or particle filter is employed to overcome this drawback since it exploits sampling techniques on non-linear observation functions.
	However, the application of Kalman filter-based methods is challenged by the following two aspects:
	First, the Kalman-filter-based methods require the noise covariance matrix as prior knowledge.
	Second, due to the matrix inversion operation in the computation of the Kalman gain, filter-based methods suffer from a high computational complexity.

	\subsubsection{Deep Learning-Based Approach} 
	With recent advancements in artificial intelligence (AI), machine learning (ML) has emerged as a promising enabling technology for rapid, low-complexity solutions in predictive beamforming.
	Specifically, ML can offload high-complexity training to the offline training stage, reducing online computational complexity. 
	In model-based implementations, the complexity-intensive steps in conventional estimation-based and non-linear filter-based approaches can be replaced by pre-trained deep neural networks, thereby leading to a reduction in computational complexity.
	This category is exemplified by Bayesian learning and deep Kalman filters.
	In model-free implementations, sequence prediction methods, such as Transformer, long short-term memory (LSTM), and gated recurrent unit (GRU), can effectively predict future mobility status based on historical observations, making them suitable for NFPB.
	
	\subsection{A Case Study of NFPB}\label{sect:case_study}
	This section proposes a case study for NFPB in a downlink single-user multiple-input single-output (SU-MISO) setup.
	For the parameter setups, the number of antennas and the carrier frequency are set to $256$ and $30$ GHz, respectively.
	Additionally, the transmit power is set to $30$ dBm, the noise power is $-50$ dBm, and the duration for one CPI is set to $10^{-2}$ s.
	We adopt the Kalman Filter-Based approach as a demonstration method.
	
	Figs. \ref{fig:trajectory} and \ref{fig:velocity} respectively show the position and velocity tracking results compared with the ground truth, where $v_x$ and $v_y$ denote the velocity components parallel to the direction of the $x$- and $y$-axis, respectively.
	It can be observed that, by harnessing the proposed NFPB, accurate and comprehensive mobility tracking is achieved in the near field without relying on any prior knowledge of the state evolution model.
	Moreover, despite the variance in the model, as indicated by the changing trajectory radius, NFPB exhibits strong generalization capability, validating the superiority discussed in Section \ref{sect:superoirity}.
	Finally, the expanded near-field region made possible by ELAAs and high carrier frequencies allows tracking over a wide area.

	\section{Research Opportunities of NFPB}
		In this section, we discuss the possible research directions for the NFPB in 6G and beyond, as shown in Fig. \ref{fig:future_directions}.
		
		\subsection{NFPB for Industrial IoT}
		In 6G and beyond, an intelligently connected industrial Internet of Things (IoT) is envisioned as a pivotal application scenario, encompassing sensing, communication, computing, and controlling.  
		However, due to the complex layout of factories, the trajectories of IoT devices are highly irregular, thus making their state evolution models challenging to construct from the viewpoint of conventional far-field sensing. 
		As a remedy, NFPB exempts the need for explicit modeling through full-dimensional mobility sensing capabilities, making it a natural fit for industrial IoT.
		Additionally, NFPB provides rich mobility status information to support intelligent control and enables low-latency delivery of control signals. 
		
		\subsection{NFPB for V2X Netwroks}
		Vehicle-to-everything communications (V2X), including vehicle-to-infrastructure (V2I) and vehicle-to-vehicle (V2V) scenarios, inherently becomes the key implementation scenario for NFPB, as its main challenge is to deal with high-mobility users under strict latency requirements.
		From the sensing perspective, the prior-knowledge-free and generalizable nature of NFPB can not only reduce the implementation cost but also system design complexity, unleashing the potential of flexible predictive beamforming.
		From the communication perspective, NFPB eliminates the dedicated feedback link between transceivers, thereby mitigating signaling delays.
		By allowing the coordination between adjacent base stations, the seamless communication experience can be fulfilled by NFPB.
		More importantly, the vehicle mobility status information offered by NFPB enables cooperative sensing and resource scheduling across the vehicular network, enhancing the overall intelligence of the transportation system.
		
		\subsection{NFPB for High-Speed Scenarios}
		For high-speed communication networks, the movements of users necessitate the design of robust communication waveforms capable of withstanding delays and Doppler frequencies, as exemplified by emerging modulation schemes including orthogonal time frequency space (OTFS), affine frequency division multiplexing (AFDM), and orthogonal delay-Doppler division multiplexing (ODDM).
		Although these waveforms provide significant performance gains in conventional far-field scenarios, their applicability to near-field conditions remains unexplored.
		The inherent limitations stem from two fundamental challenges: 1) inadequate modeling of spherical wave propagation in near-field regions, and 2) insufficient robustness against non-uniform Doppler frequencies.
		These unresolved challenges underscore the urgent need for a comprehensive investigation into next-generation waveform designs adopting more precise near-field channel models, thereby enhancing the generalizability of the waveform design to both near-field and far-field scenarios. 
		
		\subsection{NFPB for RIS-Aided Wireless Networks}
		Most existing studies on predictive beamforming focus on sensing through line-of-sight (LoS) channels.
		However, in practice, LoS conditions might be hindered when a rich-scatter environment is considered, such as urban environments.
		In such cases, the received multi-path scattering signals from non-line-of-sight (NLoS) links can originate from a large range of directions and interfere with the LoS echo signal, making location and velocity sensing more difficult.
		To address this issue, the reconfigurable intelligent surface (RIS) can be deployed to establish strong LoS links between transceivers in NLoS scenarios, thereby simplifying system design. 
		Therefore, these complex and realistic scenarios deserve more attention for the more practical implementation of NFPB.
		
		\subsection{NFPB via Flexible Antenna Designs}
		Next-generation wireless systems will feature increasingly flexible antenna architectures, opening new opportunities for NFPB. 
		In particular, the flexible antenna configurations (FAC), including fluid antennas, movable antennas, and pinching antennas, allow the antenna position to change constantly, which can facilitate the reconfiguration of small-scale or even large-scale fading.
		Given the time-varying channels induced by moving targets, NFPB can benefit significantly from the adaptability offered by FAC.
		To fully exploit the new degrees-of-freedom (DoF) of antenna systems, a key research direction lies in balancing sensing accuracy and communication performance throughout this reconfiguration process.  
		
		\subsection{NFPB for UAV/Aerial Communications}
		Unmanned Aerial Vehicles (UAVs) and high-altitude platforms represent another key domain for NFPB, owing to their inherent three-dimensional (3D) trajectories.
		By leveraging spherical waves in 3D space, NFPB enables base stations to acquire precise UAVs' full spatial information, including distance, azimuth, and elevation \cite{luo20256d}, thus achieving high-fidelity 3D localization and velocity sensing.
		
		\subsection{NFPB for PLS}
		As the confidential information is inevitably delivered by ubiquitous connectivity of 6G and beyond, physical-layer security (PLS) has garnered great attention from both academia and industry.
		Unlike conventional far-field predictive beamforming, NFPB enables full-dimensional tracking of legal user's mobility status, allowing the beam to follow legal users like ``spot lights".
		Due to the highly directional information transmission, it is more difficult for illegal users to wiretap the secret information.
		Furthermore, since NFPB exempts the need for dedicated feedback links, NFPB reduces the risk of malicious jamming attacks, further enhancing communication security.

	\section{Conclusions}
		This article proposed NFPB to enable CE-free, consistent downlink transmission for mobility communication networks.
		In particular, the fundamentals of near-field sensing and predictive beamforming were first introduced as the basics of NFPB.
		Subsequently, the NFPB was proposed, featuring its principles and advantages in comparison to the far-field counterpart.
		In addition, the implementation methods were elaborated, followed by a case study.   
		Finally, multiple research directions were pointed out to guide future research in this field.
	
	\bibliographystyle{IEEEtran}
	\bibliography{IEEEabrv,mybib}

\begin{thebibliography}{10}
\providecommand{\url}[1]{#1}
\csname url@samestyle\endcsname
\providecommand{\newblock}{\relax}
\providecommand{\bibinfo}[2]{#2}
\providecommand{\BIBentrySTDinterwordspacing}{\spaceskip=0pt\relax}
\providecommand{\BIBentryALTinterwordstretchfactor}{4}
\providecommand{\BIBentryALTinterwordspacing}{\spaceskip=\fontdimen2\font plus
\BIBentryALTinterwordstretchfactor\fontdimen3\font minus \fontdimen4\font\relax}
\providecommand{\BIBforeignlanguage}[2]{{%
\expandafter\ifx\csname l@#1\endcsname\relax
\typeout{** WARNING: IEEEtran.bst: No hyphenation pattern has been}%
\typeout{** loaded for the language `#1'. Using the pattern for}%
\typeout{** the default language instead.}%
\else
\language=\csname l@#1\endcsname
\fi
#2}}
\providecommand{\BIBdecl}{\relax}
\BIBdecl

\bibitem{liu2023survey}
Y.~Liu \emph{et~al.}, ``Near-field communications: A tutorial review,'' \emph{{IEEE} Open J. Commun. Soc.}, vol.~4, pp. 1999--2049, Aug. 2023.

\bibitem{wild2021joint}
T.~Wild \emph{et~al.}, ``Joint design of communication and sensing for beyond {5G} and {6G} systems,'' \emph{IEEE Access}, vol.~9, pp. 30\,845--30\,857, Feb. 2021.

\bibitem{liu2020radar}
F.~Liu \emph{et~al.}, ``Radar-assisted predictive beamforming for vehicular links: Communication served by sensing,'' \emph{{IEEE} Trans. Wireless Commun.}, vol.~19, no.~11, pp. 7704--7719, Aug. 2020.

\bibitem{yuan2021integrated}
W.~Yuan \emph{et~al.}, ``Integrated sensing and communication-assisted orthogonal time frequency space transmission for vehicular networks,'' \emph{IEEE J. Sel. Topics Signal Process.}, vol.~15, no.~6, pp. 1515--1528, Nov. 2021.

\bibitem{liu2022integrated}
F.~Liu \emph{et~al.}, ``Integrated sensing and communications: Toward dual-functional wireless networks for {6G} and beyond,'' \emph{IEEE J. Sel. Areas Commun.}, vol.~40, no.~6, pp. 1728--1767, Mar. 2022.

\bibitem{lu2024hierarchial}
Y.~Lu \emph{et~al.}, ``Hierarchical beam training for extremely large-scale {MIMO}: From far-field to near-field,'' \emph{IEEE Trans. Commun.}, vol.~72, no.~4, pp. 2247--2259, Apr. 2024.

\bibitem{cui2023near}
M.~Cui \emph{et~al.}, ``Near-field {MIMO} communications for {6G}: Fundamentals, challenges, potentials, and future directions,'' \emph{{IEEE} Commun. Mag.}, vol.~61, no.~1, pp. 40--46, Sept. 2023.

\bibitem{giovannetti2024performance}
C.~Giovannetti \emph{et~al.}, ``Performance bounds for velocity estimation with extremely large aperture arrays,'' \emph{IEEE Wireless Commun. Lett.}, vol.~13, no.~12, pp. 3513--3517, Oct. 2024.

\bibitem{wang2025near}
Z.~Wang \emph{et~al.}, ``Near-field velocity sensing and predictive beamforming,'' \emph{{IEEE} Trans. Veh. Technol.}, vol.~74, no.~1, pp. 1806--1810, Sept. 2025.

\bibitem{jiang2024nearfieldsensingenabledpredictive}
\BIBentryALTinterwordspacing
H.~Jiang \emph{et~al.}, ``Near-field sensing enabled predictive beamforming: From estimation to tracking,'' 2024. [Online]. Available: \url{https://arxiv.org/abs/2408.02027}
\BIBentrySTDinterwordspacing

\bibitem{yang2025near}
S.~Yang \emph{et~al.}, ``Near-field channel estimation and localization: Recent developments, cooperative integration, and future directions,'' \emph{IEEE Signal Process. Mag.}, vol.~42, no.~1, pp. 60--73, Mar. 2025.

\bibitem{zhang2022fast}
Y.~Zhang \emph{et~al.}, ``Fast near-field beam training for extremely large-scale array,'' \emph{IEEE Wireless Communications Letters}, vol.~11, no.~12, pp. 2625--2629, Oct. 2022.

\bibitem{liu2022learning}
C.~Liu \emph{et~al.}, ``Learning-based predictive beamforming for integrated sensing and communication in vehicular networks,'' \emph{{IEEE} J. Sel. Areas Commun.}, vol.~40, no.~8, pp. 2317--2334, Jun. 2022.

\bibitem{yuan2021bayesian}
W.~Yuan \emph{et~al.}, ``Bayesian predictive beamforming for vehicular networks: A low-overhead joint radar-communication approach,'' \emph{{IEEE} Trans. Wireless Commun.}, vol.~20, no.~3, pp. 1442--1456, Nov. 2021.

\bibitem{luo20256d}
H.~Luo \emph{et~al.}, ``{6D} motion parameters estimation in monostatic integrated sensing and communications system,'' \emph{IEEE Trans. Commun.}, early access, 2025, doi:{10.1109/TCOMM.2025.3534577}.

\end{thebibliography}
	
	\vspace{-1cm}
	\begin{IEEEbiographynophoto} {Hao Jiang} (Graduate Student Member, IEEE) is currently a Ph.D. student at Queen Mary University of London, U.K.  
	\end{IEEEbiographynophoto}
	
	\vspace{-1cm}
	\begin{IEEEbiographynophoto} {Zhaolin Wang} (Member, IEEE) is currently a Postdoctoral Researcher at Queen Mary University of London, U.K.  
	\end{IEEEbiographynophoto}
	\vspace{-1cm}
	
	\begin{IEEEbiographynophoto} {Yue Liu} (Member, IEEE) is currently a lecturer in the Faculty of Applied Sciences, Macao Polytechnic University, Macao SAR, China, since 2016.
	\end{IEEEbiographynophoto}
	\vspace{-1cm}
	
	\begin{IEEEbiographynophoto} {Hyundong Shin} (Fellow, IEEE) has been the Dean of College of Electronics \& Information as well as College of Software, and a Professor at Department of Electronic Engineering at Kyung Hee University.
	\end{IEEEbiographynophoto}
	
	\vspace{-1cm}
	
	\begin{IEEEbiographynophoto} {Arumugam Nallanathan} (Fellow, IEEE) is a professor and the head of the Communication Systems Research (CSR) group in Queen Mary University of London. 
	\end{IEEEbiographynophoto}
	
	\vspace{-1cm}
	\begin{IEEEbiographynophoto} {Yuanwei Liu} (Fellow, IEEE) is a (tenured) Full Professor at The University of Hong Kong and a visiting professor at Queen Mary University of London.  
	\end{IEEEbiographynophoto}
\end{document}